\documentclass[amsmath,amssymb,aps,prl,superscriptaddress,floatfix,twocolumn]{revtex4-1}

\usepackage{graphicx}
\usepackage{dcolumn}
\usepackage{bm}
\usepackage[usenames]{color}

\usepackage{hyperref} 
\usepackage{titlesec}
\usepackage{bookmark}
\usepackage{mathtools}

\newcommand{\pd}{\partial}

\renewcommand{\b}{\beta}

\renewcommand{\d}{\delta}
\newcommand{\e}{\epsilon}

\newcommand{\m}{\mu}

\newcommand{\ph}{\phi}

\newcommand{\s}{\sigma}
\renewcommand{\t}{\tau}

\newcommand{\w}{\omega}

\newcommand{\order}{\mathcal{O}}

\newcommand{\im}{\text{Im}}

\newcommand{\ra}{\rightarrow}

\newcommand{\ua}{\uparrow}
\newcommand{\da}{\downarrow}

\renewcommand{\(}{\left(}
\renewcommand{\)}{\right)}

\renewcommand{\>}{\right\rangle}

\newcommand{\lb}{\left|}
\newcommand{\rb}{\right|}

\def\beq#1\eeq {\begin{align}#1\end{align}}

\begin{document}

\title{Slope invariant $T$-linear resistivity from local self-energy}

\author{Peter Cha} 
\affiliation{Department of Physics, Cornell University, Ithaca, NY 14853, USA}

\author{Aavishkar A. Patel}
\affiliation{Department of Physics, Harvard University, Cambridge, MA 02138, USA}
\affiliation{Department of Physics, University of California Berkeley, Berkeley, CA 94720, USA}

\author{Emanuel Gull}
\affiliation{Department of Physics, University of Michigan, Ann Arbor, Michigan 48109, USA}

\author{Eun-Ah Kim}
\affiliation{Department of Physics, Cornell University, Ithaca, NY 14853, USA}

\date{\today}

\begin{abstract}

A theoretical understanding of the enigmatic linear-in-temperature ($T$) resistivity, ubiquitous in strongly correlated metallic systems, has been a long sought-after goal. Furthermore, the slope of this robust $T$-linear resistivity is also observed to stay constant through crossovers between different temperature regimes: a phenomenon we dub ``slope invariance''.  
Recently, several solvable models with $T$-linear resistivity have been proposed, putting us in an opportune moment to compare their inner workings in various explicit calculations. We consider two strongly correlated models with local self-energies that demonstrate $T$-linearity: a lattice of coupled Sachdev-Ye-Kitaev (SYK) models
and the Hubbard model in single-site dynamical mean-field theory (DMFT).  We find that the two models achieve $T$-linearity through distinct mechanisms at intermediate temperatures. However, we also find that these mechanisms converge to an identical form at high temperatures. Surprisingly, both models exhibit ``slope invariance'' across the two temperature regimes. 
We thus not only reveal some of the diversity in the theoretical inner workings that can lead to $T$-linear resistivity, but we also establish that different mechanisms can result in ``slope invarance''.
\end{abstract}

\maketitle

The mysterious incoherent metallic states with $T$-linear resistivity ($\rho_{\rm DC}\propto T$) seen in many strongly correlated materials \cite{Martin1990, Hussey2009, Mackenzie2013} have long puzzled researchers, as such a temperature dependence is inaccessible from Fermi liquid theory \cite{Kivelson-qp}. Particularly remarkable is the fact that the slope $d\rho_{\rm DC}/dT$ remains constant as $T$ varies over $2$ or $3$ orders of magnitude, while the temperature shoots through multiple crossover energy scales: we dub this phenomenon ``slope invariance''. Recent progress \cite{balents,patel,senthilsyk,syk-sc,hartnoll,PS2019} in solvable strongly interacting models that yield $T$-linear resistivity  and through computational and experimental quantum simulation~\cite{coldatom} have injected renewed enthusiasm and hope into the community. Given that ``solvability'' requires unrealistic limits however, theoretical insight into the common or unique mechanisms of the $T$-linear resistivity obtained in these models is much needed in making contact with experimental observations. 

 The unusual feature of local self-energy is at play in two of the most studied microscopic models with $T$-linear resistivity: lattice models of coupled SYK quantum dots~\cite{balents,patel,senthilsyk,syk-sc} and its earlier incarnation of a doped random-bond Heisenberg model~\cite{parcollet}; and the Hubbard model in single-site dynamical mean-field theory (DMFT)~\cite{coldatom, badmetalsgood, kokalj}. Moreover, in both models, exact non-trivial self-consistency equations enable an explicit computation. Yet no comparative study has been carried out. In fact, these models have remained largely only of theoretical interest because $T$-linear resistivity in these models is only reached at temperatures above certain crossover scales that are too high for solid-state systems. However, in a recent emulation of the Hubbard model using ultracold atomic gases~\cite{coldatom}, $T$-linear resistivity was observed at ``intermediate temperatures" ($t<T<U$,  where $t$ and $U$ are the hopping and interaction energies respectively) accessible in model calculations. Furthermore, ``slope invariance'' is also evident in the data.  This motivates a fresh consideration of solvable models with $T$-linearity at intermediate- and high-temperatures seeking an insight into slope invariance. 
 In this letter we compare the inner workings of transport in two models with local self-energy, coupled SYK and DMFT, in two temperature ranges: intermediate ($t\ll T\ll U$) and high temperature ($t\ll U\ll T$). We find distinct mechanisms at intermediate temperatures converging to a unifying picture at high temperature. Further we note how the unifying picture ties to ``slope invariance''.

For both models of interest, vertex corrections to the conductivity vanish~\cite{patel,senthilsyk, dmftrev}. Hence the DC conductivity can be compactly written in the spectral representation of the Kubo formula
\beq
\sigma_{\rm DC}=&2\pi\int d\e\ \ph(\e)\int \frac{\b d\w\(A(\e,\w)\)^2}{4\cosh^2(\b\w/2)},\label{Kubo}
\eeq
where $\phi(\e)=\sum\limits_{\mathbf{k}}\(\frac{\pd \e(\mathbf{k})}{\pd k_x}\)^2\d(\e-\e(\mathbf{k}))$ is the transport function and $A(\e,\w)=-\frac{1}{\pi}{\rm Im}G(\e,\w)$ is the spectral function, normalized to $\int d\w A(\e,\w)=1$. Hence different mechanisms of $T$-linear resistivity arise from different functional forms of the spectral function. 

{\it Intermediate Temperatures --} We first review recent results finding $T$-linear resistivity in coupled SYK models~\cite{patel, senthilsyk,syk-sc, balents}.
These models are the best understood among the solvable models with $T$-linear resistivity and they may be realizable in multi-orbital systems~\cite{Werner2018, Tsuji2019}. In the SYK models,
disorder averaging enforces the constraint that the self-energy of conduction electrons is local $\Sigma(\vec{\mathbf{k}},i\omega)\equiv\Sigma(i\omega)$, as long as the coupling between SYK dots is perturbative. The self-consistent Dyson equations then arise in closed-form as the saddle-point equations of a large-$N$ limit (where each SYK dot consists of $N$ flavors of fermions), in which the self-energy diagrams are truncated at 2nd order. At intermediate and high temperatures, the hopping between lattice sites is perturbative, so that the lattice Green's function is also local $G(\vec{\mathbf{k}}, i\omega)\equiv G(i\omega)$ and the Dyson equations take the general form
\beq
&G(i\w)^{-1} = i\omega + \mu-\Sigma(i\w),\nonumber \\
&\Sigma(\t) = -U^2G(\t)^{q/2}G(-\t)^{q/2-1}.
\label{SYKDyson}
\eeq
Here $q$ is the fermionic degree of the disordered SYK interaction in the Hamiltonian and $\mu$ is the chemical potential. In the intermediate-temperature regime, where $T\ll U$, the $i\omega$ term in the first line can be ignored. At half-filling, these equations then have the solution $G(\w)\propto-i\b^{1-2/q}g_q(\b\w)$,
where we have defined for convenience $g_q(x)=\Gamma(1/q-ix/2\pi)/\Gamma(1-1/q-ix/2\pi)$.

Due to the local spectral function, the Kubo expression (\ref{Kubo}) then further simplies to $\sigma_{\rm DC}\propto\int \frac{\b d\w\(A(\w)\)^2}{4\cosh^2(\b\w/2)}$. Hence, 
\beq
\sigma_{\rm DC}\propto& t^2\b^{2(1-2/q)}\int dx \frac{(g_{q}(x))^2}{\cosh^2(x/2)}
\propto t^2\b^{2(1-2/q)},\label{SYKcond}
\eeq
given the solution to the Dyson equation. This streamlined derivation makes clear that $T$-linear resistivity requires $q=4$ \cite{senthilsyk, Note1}, in which case the complete expression for the conductivity at intermediate temperatures $t^2/U \ll T \ll U$ is
\beq
\sigma_{\rm DC}=\frac{t^2\b\sqrt{\pi}}{2 U \sqrt{\cosh 2\pi\mathcal{E}}},
\label{SYKcold}
\eeq
\cite{balents,patel} where $\mathcal{E}$ is a function solely of the fermion filling that vanishes at half-filling~\cite{Davison2017, Note2}. Two features of this calculation offer clarifying insight: First, the SYK spectral function is a scaling function of the dimensionless parameter $\b\w$, and extends broadly over the entire frequency range below the UV cutoff $|\w|< U$. (Appendix B) Second, although local self-energy is a generic feature of SYK models, $T$-linear resistivity in the intermediate temperature range is a consequence of the particular scaling exponent of the local Green's function within the ``quantum dots'' for $q=4$. The same mechanism was found earlier in a doped random Heisenberg model in Ref.~\cite{parcollet}.

\footnotetext[1]{$T$-linearity can be achieved for other values of $q$ by changing the fermionic degree of the hopping term \cite{syk-sc}. For example, with $p$-fermion hopping
\beq
\sum t_{i_1\ldots j_p}c^\dag_{i_1}\ldots c^\dag_{i_{p}}c_{j_{1}}\ldots c_{j_p},
\eeq
the expression for conductivity is modified as
\beq
\sigma_{\rm DC}=&2\pi\int d\w d\w_1\ldots d\w_{2p-2}\frac{\b A(\w)^{2p}}{\cosh^2(\b\w/2)}\\
\propto& \b^{2(1-2p/q)},
\eeq
showing that $T$-linearity is achieved when $q=4p$.}

\footnotetext[2]{The conductivity crosses over from (\ref{SYKcold}) to Fermi-liquid like $1/T^2$ behavior in the ``low-temperature'' regime of $T\ll t^2/J$~\cite{parcollet, balents,senthilsyk}. However, we always focus on a higher temperature range.}

Now we turn to the DMFT study of the repulsive $U$ Hubbard model on the two-dimensional square lattice:
\beq
H=-t\sum_{\<i,j\>,\s}c^\dag_{i\s}c_{j\s}+U\sum_{i}n_{i\ua}n_{i\da}-\m\sum_{i\s}n_{i\s}.\label{eq:hubbard}
\eeq
In DMFT, we first integrate out all sites except one to reduce the problem to a local impurity model. Then the infinite-dimension approximation imposes a self-consistency constraint where the bare propagator of the impurity model is determined by the fully dressed 2-point function computed within the impurity model. Lacking an additional large-$N$ parameter of SYK, a closed-form Dyson equation is not available, however. Hence the  impurity model is solved computationally.
Despite intense efforts at computational studies of the model and its extensions and observations of $T$-linear resistivity upon certain approximations~\cite{badmetalsgood, devereaux,kokalj,vucicevic2015, vucicevic2019}, new interest in models with local self-energies warrant a close contextual look at the DMFT solutions to the Hubbard model. 
Here we carry out single-site DMFT  using an exact diagonalization (ED) impurity solver that frees us from the need of analytic continuation.

\begin{figure}
\includegraphics[width=3in, trim={0 0 0 1cm},clip]{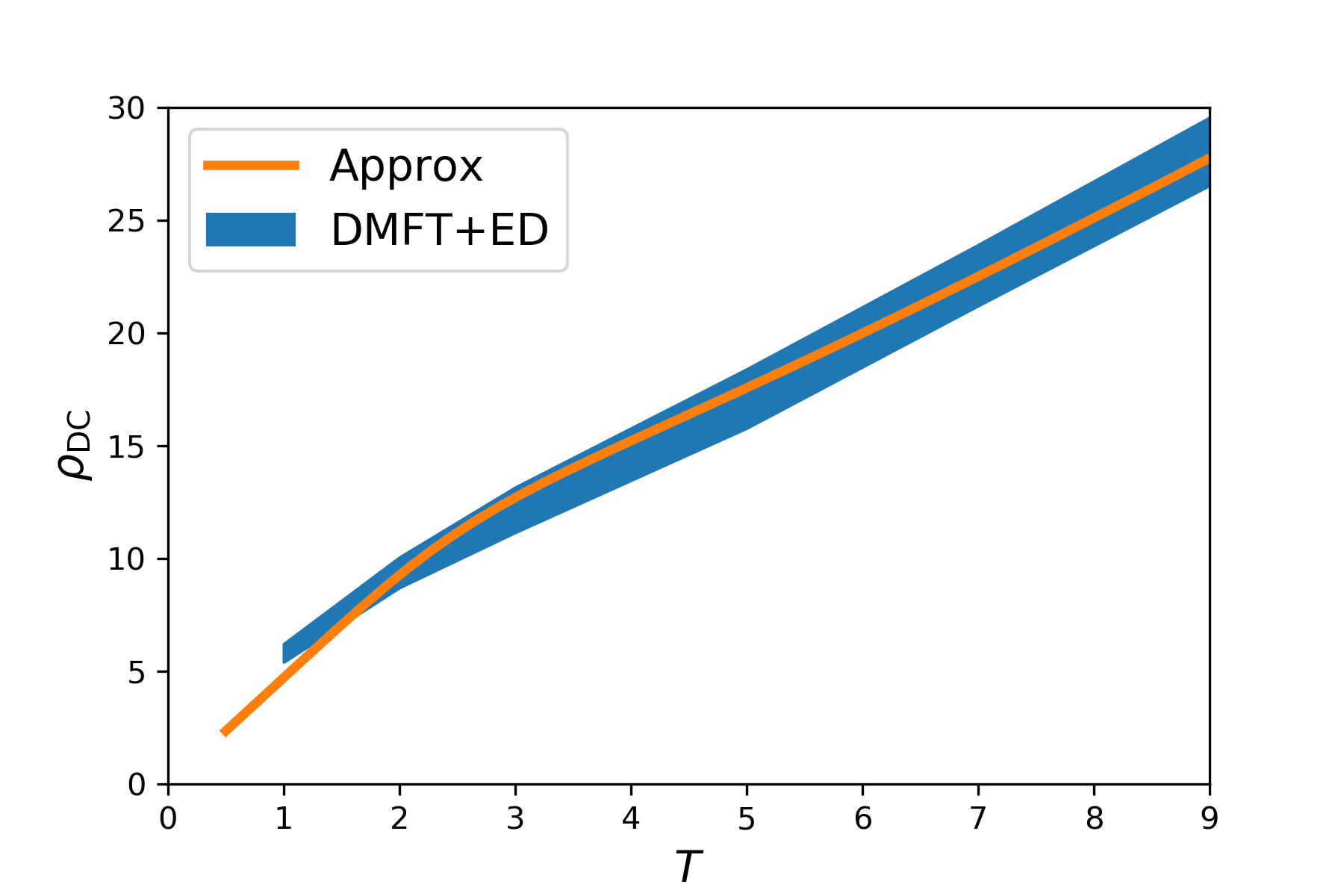}
\caption{Resistivity $\rho_{
\rm DC}$ as a function of Temperature, in single-site DMFT with $U=12$ at electron filling $n=0.825$, using an ED impurity solver with $n_s=8$ total sites. Vertical spread of blue curve comes from varying the ED broadening parameter $\eta=0.01\sim0.08$. Orange curve displays the approximation scheme of Eq.~(\ref{eq:DMFT-condapprox}).
\label{fig:resistivity}}
\end{figure}

We explored the range of $7.5<U<12$ and $0.2<T<9$ in units of hopping $t$, with the electron density per site $n=0.825$.
In Fig. \ref{fig:resistivity}, we plot a representative result. 
Here the width of the band represents the dependence of $\rho_{\rm DC}(T)$ on the ED broadening parameter $\eta$.
The resulting curve is clearly linear in the intermediate temperature range $t\ll T\ll U$ as well as high temperature with a small slope change in between as has been seen earlier~\cite{Note6}.

\footnotetext[6]{Similar results are seen in the DMFT calculation of resistivity in Brown et al \cite{coldatom}, with the parameters $n=0.825$, $U=7.5$, $T<6$, with $n_s=6$.}

\begin{figure}
\includegraphics[width=3in, trim={0 0 0 1.2cm},clip]{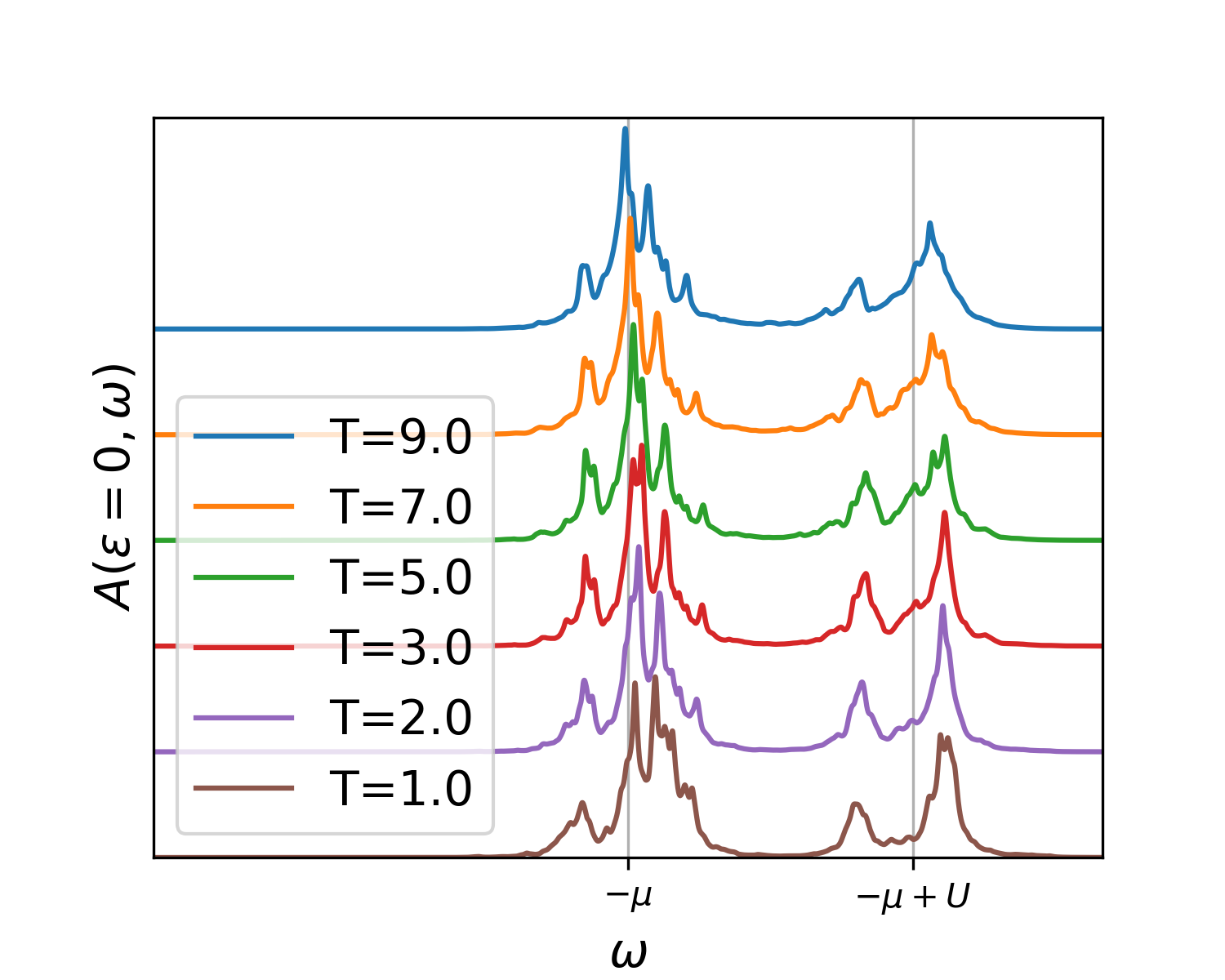}
\caption{$A(\e=0, \w)$, from DMFT with $U=12$ at electron filling $n=0.825$, using ED impurity solver with $n_s=8$ and $\eta=0.08$. Frequencies have been shifted by the chemical potential. \label{fig:imgreen}}
\end{figure}

In search of analytic insight into $T$-linear resistivity at intermediate temperatures $t\ll T\ll U$, 
we examine the DMFT spectral function $A(\e,\w)$ in detail for a suitable analytic ansatz. 
The DMFT spectral function is largely $T$-independent at the intermediate temperatures being shown, and consists of a lower and an upper band of widths $\sim t$ that are separated in frequency by $\approx U$ (see  Fig.~\ref{fig:imgreen}).
Hence we approximate the DMFT spectral function using the following ansatz:
\beq
A(\e, \w)=a_l h(\w; \w_l,\eta_l)+a_u h(\w;\w_u, \eta_u),\label{eq:DMFT-Aapprox}
\eeq
where the lower and upper bands $A_{l,u}(\e,\w)$ have weights $a_{l,u}$, satisfying $a_l+a_u=1$, and are localized in frequencies at $\w_{l,u}$ with widths $\eta_{l,u}\ll T$. For simplicity, we model each band $h(\w)$ by a normalized Gaussian, centered at $\w_{l,u}$ with standard deviation $\eta_{l,u}$.
Then the parameters for the lower band $a_l$, $\w_l$, $\eta_l$ are fit to the DMFT spectral function by $\int d\w\ A_l(\e,\w)=a_l$, $\int d\w\ \w A_l(\e,\w)=a_lw_l$, and $\int d\w\ (A_l(\e,\w))^2=\(1/2\sqrt{\pi}\)a_l^2/\eta_l$, and likewise for the upper band parameters \cite{Note3}. The parameters depend on $\e$ but are independent of $T$.
\footnotetext[3]{We have selected a Gaussian form for $h(\w)$ in the text, but we note that the precise forms of each band are not crucial to the results. Using a different bounded function for $h(\w)$ (e.g. Lorentzian) will only affect the proportionality constant in front of $a^2/\eta$ in the final fitting equation.}
With this ansatz, we arrive at a simplified explicit expression for DC conductivity (see Appendix A for a detailed derivation)
\beq
\sigma_{\rm DC}=\b&\(\frac{C_l}{4\cosh^2(\b\m/2)}+\frac{C_u}{4\cosh^2(\b(U-\m)/2)}\)\label{eq:DMFT-condapprox}
\eeq
where $C_{l,u}\equiv\sqrt{\pi}\int d\e\ \phi(\e)a_{l,u}^2/\eta_{l,u}$. 
In the intermediate temperature range, Eq.~\eqref{eq:DMFT-condapprox} yields $T$-linear resistivity by the first term in the paranthesis approaching a constant ($\mu\propto T$) and second term vanishing ($T\ll U$). We note that similar approximations were employed in previous works~\cite{Palsson1998, mukerjee2007}, which yielded similar expressions for conductivity.

Fig.~\ref{fig:resistivity} demonstrates that the ansatz Eq.~\eqref{eq:DMFT-condapprox} (shown in orange) captures the DMFT results on resistivity excellently over the whole temperature range. This fit reveals that key mechanism for $T$-linear resistivity in DMFT is the 
broadening of the spectral function by a $T$-independent scattering rate $1/\tau$ controlled by a hopping scale $\sim t$. Such a highly localized spectral function leads to $\s_{\rm DC}\propto \tau/T$ from a linearization of the Fermi factor in the Kubo formula. 
We suspect the same mechanism is at play for $t\ll U$ modified Hubbard model studied by Mousatov et al \cite{hartnoll} where the broadening in that case is set by a longer ranged repulsion $V$.
This is in contrast to the inner workings of coupled SYK, where the spectral function is a broad, singly-peaked scaling function of $\b\w$ and $T$-linear resistivity is specific to $q=4$. (Appendix B)

Armed with the above analytic insight, we now turn to the Lorenz ratio $L=\kappa/\sigma T$. The Wiedemann-Franz law with $L_{WF}=\pi^2/3$ is a well-known property of a Fermi liquid with elastic scttering. Although it can be violated even in a Fermi liquid in the presence of inelastic scattering  \cite{taillefer1997,pourovskii2017,WF-FL}, it is nevertheless a useful quantity to evaluate. 
The thermal and electrical conductivities can be expressed in terms of kinetic coefficients
\beq
L_n=&2\pi\int d\e\ \ph(\e)\int \frac{\b d\w (\b\w)^n \(A(\e,\w)\)^2}{4\cosh^2(\b\w/2)}
\eeq
as $\sigma = L_0$ and $\kappa = T(L_2-(L_1)^2/L_0)$. Once again employing our approximate spectral function Eq.~\eqref{eq:DMFT-Aapprox}, we find the violation of the Wiedemann-Franz law with the Lorenz ratio vanishing as a power-law in temperature for $t \ll T$.
In contrast, the coupled SYK model of \cite{balents} violates the Wiedemann-Franz law only through modification of the Lorenz ratio $L = 0.375\times L_{WF}$ \cite{balents,patel} in the intermediate temperature regime with $T$-linear resistivity.

\begin{figure}
\includegraphics[width=3in]{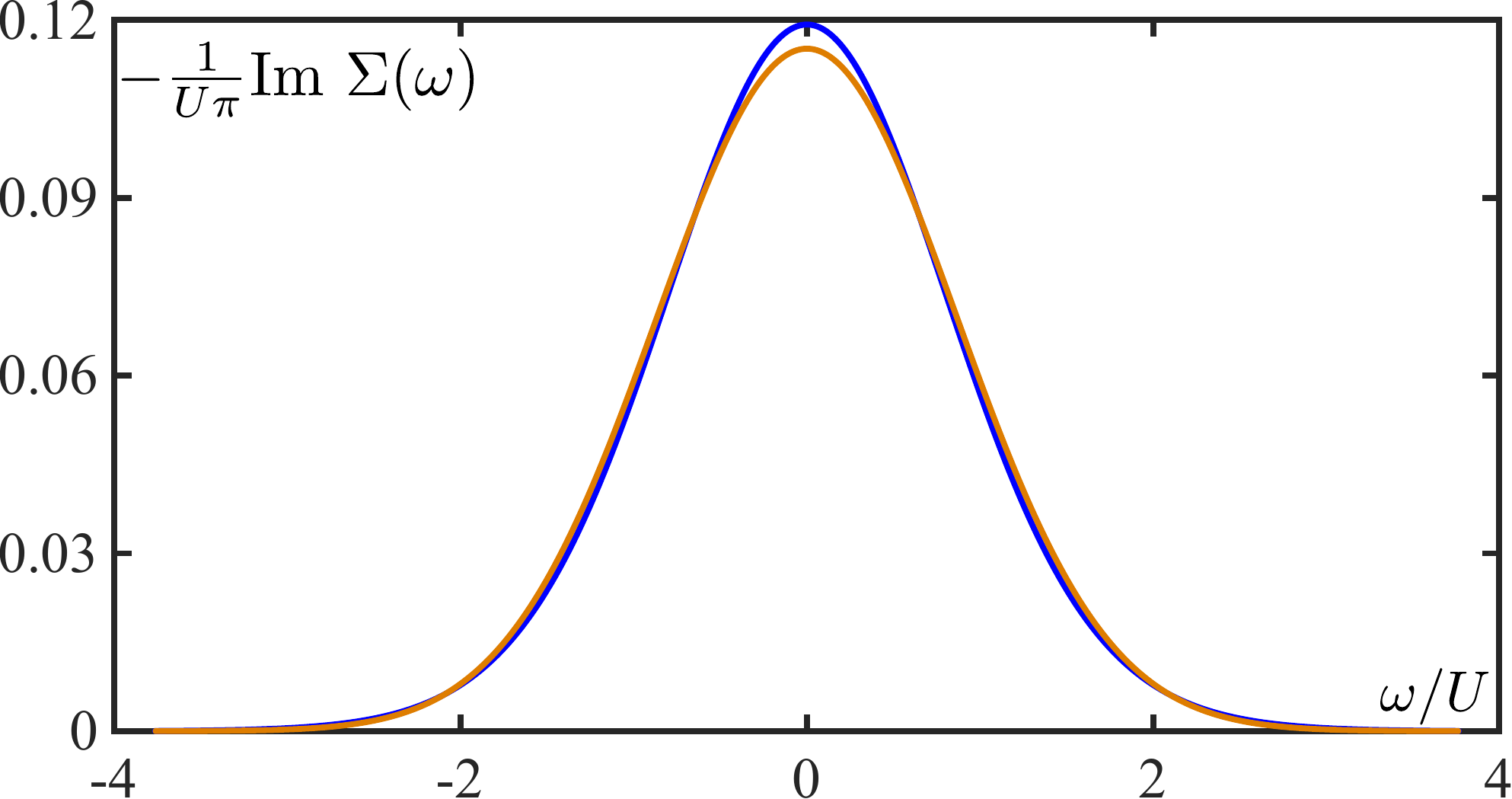}
\caption{Comparision of the numerically obtained self-energy (blue) in for the $q=4$ SYK model in the high-temperature regime with a Gaussian ansatz (orange) for analytically continuing a high-temperature expansion (\ref{SYKGaussian}). Here $\beta U = 0.01$, and the system is at half-filling.}
\label{SYKhot}
\end{figure}

{\it High Temperature Limit --} We now consider the high-temperature regime, where $T$ is the largest scale in the problem. Resistivity at high-temperature is often overlooked, as all Hamiltonians with a bounded spectrum of states are known to have $T$-linear resistivity in this temperature regime~\cite{Note4, hightemp}. This argument is useful from a formal perspective but offers little insight into $T$-linear resistivity at intermediate temperatures, or the lack of slope change across the crossover $T\sim U$. In this ``weak coupling'' limit, we approximate single-site DMFT using self-consistent second-order perturbation theory (GF2)~\cite{Phillips14, Iskakov18, Motta17, Rusakov18}. Unlike bare second order perturbation theory, GF2 is $\Phi$-derivable~\cite{Luttinger60} and therefore thermodynamically consistent and symmetry conserving~\cite{BaymKadanoff61,Baym62}, implying that thermodynamic relations and conservation laws are intrinsically satisfied. We find, within this high-temperature limit, that the DMFT equations converge to the self-consistency equations of $q=4$ SYK!

To see the convergence note that  the self-energy $\Sigma(i\w)$ is the sum of a tadpole diagram and a sunset diagram in GF2, i.e.,   $\Sigma(i\w)=\Sigma^{(1)}(i\w)+\Sigma^{(2)}(i\w)$, where $\Sigma^{(1)}(i\w)=UG(\t=\b^-)=Un/2$ and $\Sigma^{(2)}(\tau)=-U^2G(\t)^2G(-\t)$. Now the Dyson equations take a closed form: 
\beq
G(i\w)^{-1}=&i\w+\mu_{\mathrm{eff}}-\Sigma^{(2)}(i\w),\nonumber\\ 
\Sigma^{(2)}(\tau)=&-U^2G(\t)^2G(-\t)\label{2PTDyson},
\eeq
which are identical to the $q=4$ SYK equations~\eqref{SYKDyson}, with a shifted chemical potential $\mu_{\mathrm{eff}}=\mu-nU/2$. This discovery allows us to simultaneously treat single-site DMFT and $q=4$ SYK.

\footnotetext[4]{Consider the spectral representation of conductivity $\s(\w)=-\frac{\pi}{\w}\sum\limits_{nm}\lb< n\lb J_x\rb m>\rb^2e^{-\b E_n}\(e^{\b E_{nm}}-1\)\d(\w-E_{nm})$ where $n,m$ labels the many-body eigenstates and $E_{nm}=E_n-E_m$. When $E_n\ll T$ for all $n$, the Boltzmann factors can be expanded to give a single factor of $\b$ in the expression to leading order, which shows $T$-linear resistivity.}

Interestingly, with $T\gg U$, we find, by numerical solution of the real-time version of~\eqref{2PTDyson}, that the self-energy is temperature-independent and Gaussian to a very good approximation (Fig.~\ref{SYKhot}). By using a high-temperature expansion in imaginary-time~\cite{Fu2016} and the ``maximum entropy" ansatz for analytic continuation~\cite{hightemp}, we further determine, for $\m_{\text{eff}}=0$,
\beq
-\frac{1}{\pi}\im~\Sigma(\omega) \approx \frac{U}{2\pi}\sqrt{\frac{\pi}{6}}\mathrm{exp}\left(-\frac{2\omega^2}{3U^2}\right).
\label{SYKGaussian}
\eeq
For $\mu_{\text{eff}}\neq0$ we find a similar function with its peak shifted to $\omega=\m_{\text{eff}}$.
A Gaussian self-energy of width $\sim U$ leads to a spectral function also of width $\sim U$. In the high temperature regime $U\ll T$, this leads to $T$-linear resistivity in the same manner as~\eqref{eq:DMFT-condapprox}, but with only one band at $-\mu_{\text{eff}}$.

\begin{figure}[t]
\includegraphics[width=3in, trim={0 0 0 1.2cm},clip]{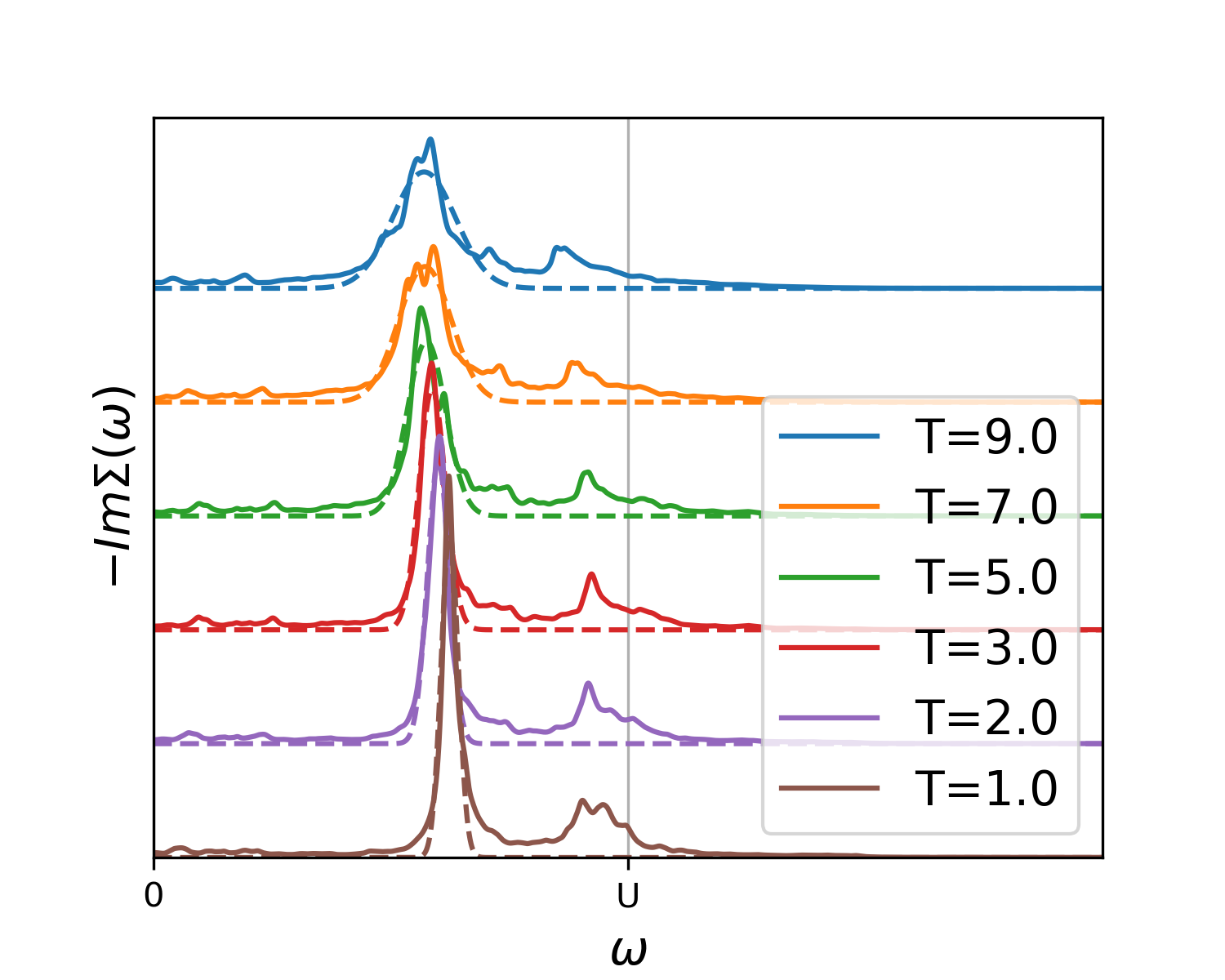}
\caption{Imaginary self-energy $-\im~\Sigma(\w)$ vs real frequency $\w$, from DMFT with $U=12$ at electron filling $n=0.825$, using ED impurity solver with $n_s=8$ and $\eta=0.08$. Dashed line plots a Gaussian fit to the central peak of self-energy at each temperature. Frequencies have been shifted by $\mu_{\text{eff}}$. \label{fig:imsig}}
\end{figure}

\begin{figure}
\includegraphics[width=3in]{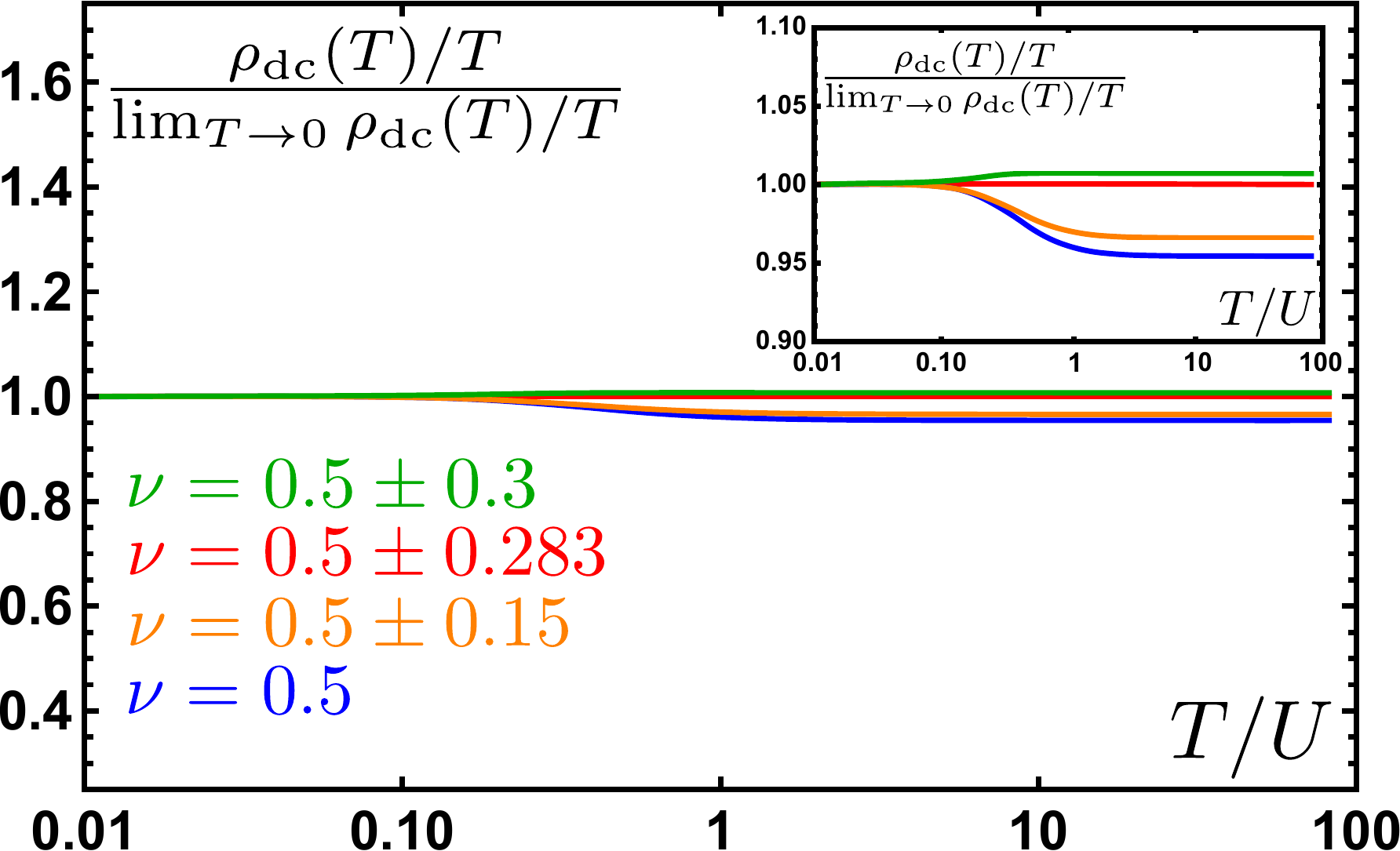}
\caption{Normalized change in slope of linear-in-$T$ resistivity for the $q=4$ SYK chain with infinitesimally weak quadratic hopping at different fillings $\nu$. The inset shows the same plot with the y-axis magnified.}
\label{SYKslope}
\end{figure}

The divergence of the two models at intermediate temperatures is visible in how the asymptotic form of self-energy responds to the lowering of temperature. In DMFT, the self-energy 
becomes increasingly sharply peaked at lower temperatures. This trend is visible in Fig.~\ref{fig:imsig}, which displays the self-energy in DMFT across a range of intermediate temperatures for $U=12$ and $n=0.825$. We have also plotted a Gaussian fit to the primary peak in dashed lines in Fig.~\ref{fig:imsig}.  Gaussian self-energy in DMFT at intermediate temperatures was suggested in \cite{hightemp}, by Maximum Entropy analytic continuation. In sharp contrast, the self-energy in $q=4$ coupled SYK undergoes a dramatic change in form, departing from the Gaussian form at high-temperatures to  $-\im~\Sigma(\omega\ll U)\sim \max(\sqrt{UT}, \sqrt{U\omega})$. (Appendix B) Surprisingly, despite this qualitative shift in the single-particle properties, the slope change in resistivity $d\rho_\text{DC}/dT$ remains negligible across $T\sim U$ for all numerically accessible values of the filling~\cite{Note5}, and actually goes to zero within numerical tolerances at fillings $\approx 0.5 \pm 0.283$ (Fig.~\ref{SYKslope}). 

The contrasting forms of the DMFT and SYK self-energies below the high-temperature limit can be traced to the validity of GF2.
In the SYK incoherent metal, GF2 is exact independently of temperature because the large number of degrees of freedom $N$, with only disordered interactions among them, explicitly makes all self-energy diagrams beyond second order non-thermodynamic.
This exact truncation cannot be done in single-site DMFT where each local electron degree of freedom has only self-interactions and there is no such large-$N$ limit.
Thus GF2 is approximately valid in DMFT only in the high-temperature limit where interactions are effectively perturbative. As temperature is lowered across the crossover scale $U$, higher order processes of $\order((U/T)^3)$ become relevant and lead to a breakdown of GF2.

{\it Discussion -- }  At intermediate temperatures $T\ll U$, we observed that single-site DMFT achieves $T$-linear resistivity by grouping the spectrum of states into two narrow bands of widths far below the interaction scale $U$. In the same temperature regime, SYK lattice models have a broad spectrum that extends over the entire range $|\w|\lesssim U$, and give $T$-linear resistivity for models with $q=4$, for which $A(\w)\sim 1/\sqrt{U\w}$ for $T\ll \w\ll U$. Although the intermediate-temperature mechanisms behind $T$-linear resistivity in these models are seemingly unrelated, we have made the surprising observation that the models seem to converge to an identical form at high temperature $T\gg U$.
We further observe that the $q=4$ SYK model shows almost no slope change in its $T$-linear resistivity across the crossover scale $T\sim U$ (Fig.~\ref{SYKslope}).

\footnotetext[5]{For fillings $\gtrsim 0.81$ and $\lesssim 0.19$, the $q=4$ SYK model has a first-order phase transition to a nearly-free fermion phase at temperatures $T\ll U$~\cite{Azeyanagi2018}, and we are hence unable to explore those regions numerically.}

Our finding of ``slope invariance'' in lattice SYK despite major changes in the single-particle spectral properties is remarkably reminiscent of observations made in strongly correlated materials. On the other hand, the slope in the two temperature regimes have distinct doping dependences in DMFT and slope invariance requires fine tuning in doping \cite{hightemp}. However, the cross-over temperature is suppressed from the na\"ive $\sim U$ to $\sim U/\log(1/\d^2)$ at small dopings \cite{hightemp} due to temperature dependence of chemical potential. This accounts for the observation that the change in slope in Fig. \ref{fig:resistivity} occurs at $T\approx 2=U/6$, as opposed to at $T\approx U$. Furthermore, the slopes in the two temperature regimes differ by less than $11\%$ for all large enough doping $d>0.33$. (See Appendix C) The renomalized cross-over scale can push the slope change outside of range of observation which may explain the featureless $T$-linear resistivity reported in Ref.\cite{hartnoll}. Finally, the change in $T$-dependence of chemical potential that drives the slope change in $T$-linear resistivity is also responsible for a slope change in inverse compressibility $\chi^{-1}$ (see Appendix C), which can further complicate any inference on diffusivity from the Nernst-Einstein relation $\sigma_{\rm DC}=\chi \mathcal{D}$ and calculation of $\chi$ \cite{calandra2003, hartnoll, devereaux}.

\section{Acknowledgments}\label{sec:ack}

We thank Erez Berg, Antoine Georges, David Huse, Srinivas Raghu and Subir Sachdev for helpful discussions. PC and E-AK were supported by the W.M. Keck Foundation. AAP was supported by the US Department of Energy under Grant No. DE-SC0019030, and by the Miller Institute for Basic Research in Science. EG acknowledges sabbatical support by the Center for Computational Quantum Phsyics of the Flatiron Institute. This work was initiated at Aspen Center for Physics, which is supported by National Science Foundation grant PHY-1607611.

\bibliographystyle{apsrev4-1}
\bibliography{biblio}

\clearpage

\section{Appendix}

\subsection{A. Detailed derivation of conductivity}

In this section we derive the expression for conductivity \eqref{eq:DMFT-condapprox} from the Kubo formula 
\beq
\sigma_{\rm DC}=&2\pi\int d\e\ \ph(\e)\int \frac{\b d\w\(A(\e,\w)\)^2}{4\cosh^2(\b\w/2)}.
\eeq
From the DMFT spectral function in Fig.~\ref{fig:imgreen}, we employ the ansatz of \eqref{eq:DMFT-Aapprox}
\beq
A(\e, \w)=a_l h(\w; \w_l,\eta_l)+a_u h(\w;\w_u, \eta_u),
\eeq
where the lower and upper bands $A_{l,u}(\e,\w)$ have weights $a_{l,u}$, satisfying $a_l+a_u=1$, and are localized in frequencies at $\w_{l,u}$ with widths $\eta_{l,u}\ll T$. Note that the ansatz parameters $a_{l,u}$, $\w_{l,u}$, and $\eta_{l,u}$ can depend on the band energy $\e$.

We can now scale out $\b$ by changing the variable of integration $x\equiv \b\w$. The spectral function in terms of $x$ is
\beq
A(\e, x)=\beta\(a_l h_x(x; \beta\w_l,\beta\eta_l)+a_u h_x(x;\beta\w_u, \beta\eta_u)\).
\eeq
If the spectral function consists of widely separated narrow peaks $\eta_{l,u}\ll|\w_u-\w_l|$, then inter-band processes will be suppressed
\beq
A(\e, x)^2\approx\beta^2\(a_l^2 h_x(x; \beta\w_l,\beta\eta_l)^2+a_u^2 h_x(x;\beta\w_u, \beta\eta_u)^2\).
\eeq
The $x$-integral can now be carried out explicitly. We focus on the lower band for brevity but the calculation for the upper band is identical.
\beq
\int \frac{\b d\w\(A(\e,\w)\)^2}{4\cosh^2(\b\w/2)}=&\int \frac{dx\ \beta^2a_l^2 h_x(x; \beta\w_l,\beta\eta_l)^2}{4\cosh^2(x/2)}\\
=&\frac{\b^2 a_l^2}{4\cosh^2(\beta\w_l/2)}\frac{1}{2\sqrt{\pi}\beta\eta_l}\\
=&\frac{1}{2\sqrt{\pi}}\frac{\b a_l^2}{4\eta_l\cosh^2(\beta\w_l/2)}\\
\approx&\frac{1}{2\sqrt{\pi}}\frac{\b a_l^2}{4\eta_l\cosh^2(\beta\m/2)}\\
\eeq
where in the final line we make the observation that $|\w_l+\m|\ll T$ for all $\e$. The lower band contribution to the conductivity can thus be written
\beq
\sigma_{{\rm DC},l}=&\b\frac{C_l}{4\cosh^2(\b\m/2)}
\eeq
where we have defined $C_l\equiv\sqrt{\pi}\int d\e\ \phi(\e)a_{l,u}^2/\eta_{l,u}$. Summing the contribution from the upper band, we arrive at the expression for conductivity \eqref{eq:DMFT-condapprox}.

\subsection{B. Comparative Plot}
\begin{figure*}
\includegraphics[width=6in]{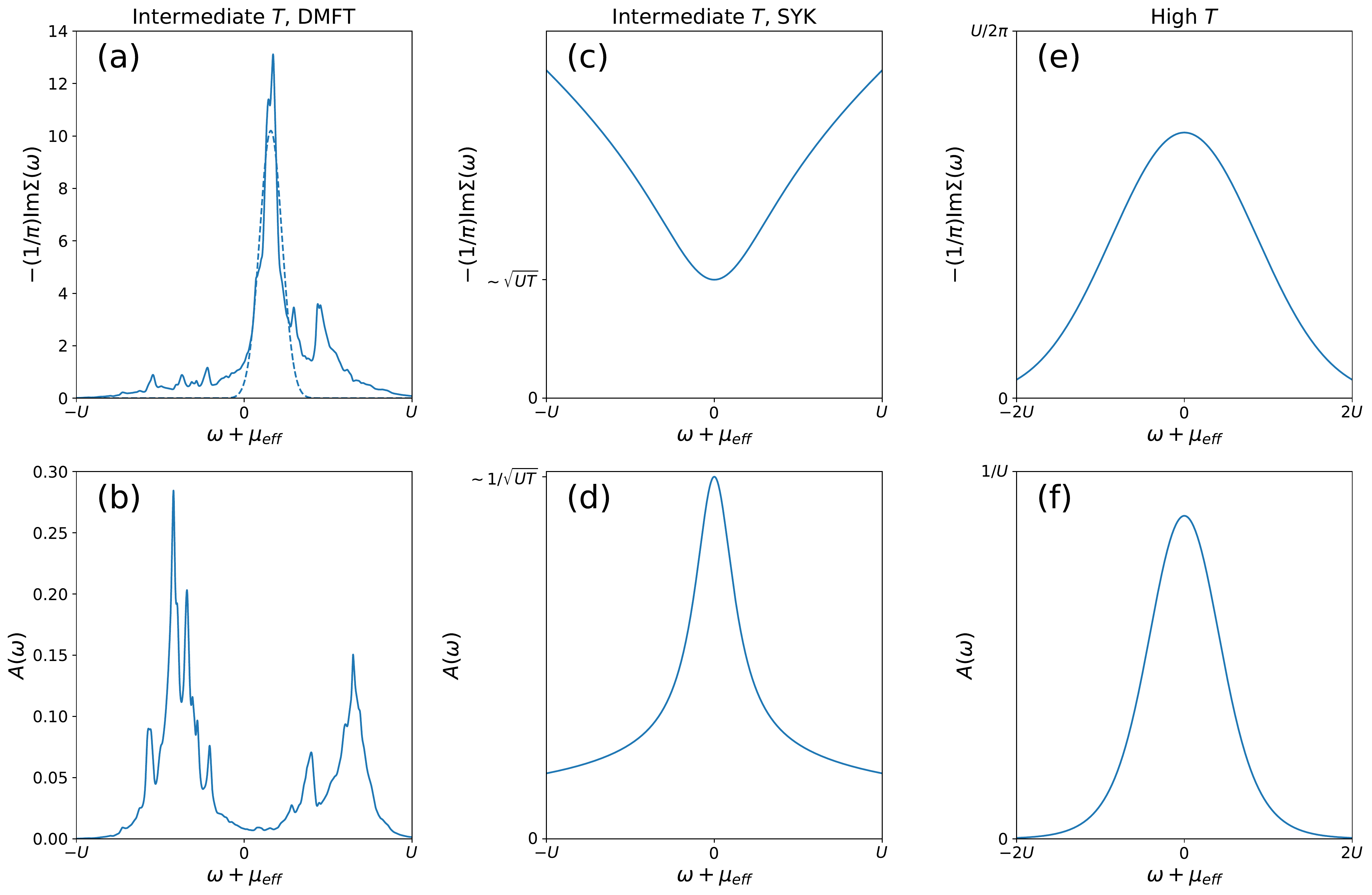}
\caption{\label{fig:compplot} 
Plots of self-energy (a, c, e) and spectral function (b, d, f) vs frequency shifted by $\mu_{eff}$. The dashed line in (a) displays the gaussian fit of Fig.~\ref{fig:imsig}. The singly-peaked self-energy of DMFT at intermediate $T$ (a) contrasts with the self-energy of SYK also at intermediate $T$ (c), which has a local minimum at $\omega+\mu_{eff}=0$, while the DMFT spectral function (b), which is highly localized in freqency, contrasts with the broad scaling form of the SYK spectral function (d). Despite these contrasting forms at intermediate temperature, the two models converge to the Gaussian self-energy form at high temperature (e, f).} 
\end{figure*}

In Fig.~\ref{fig:compplot} we show typical plots of the self-energies and spectral functions in the two temperature regimes from the DMFT calculation and the coupled SYK calculation.

\subsection{C. Compressibility}

\begin{figure}
\includegraphics[width=3in]{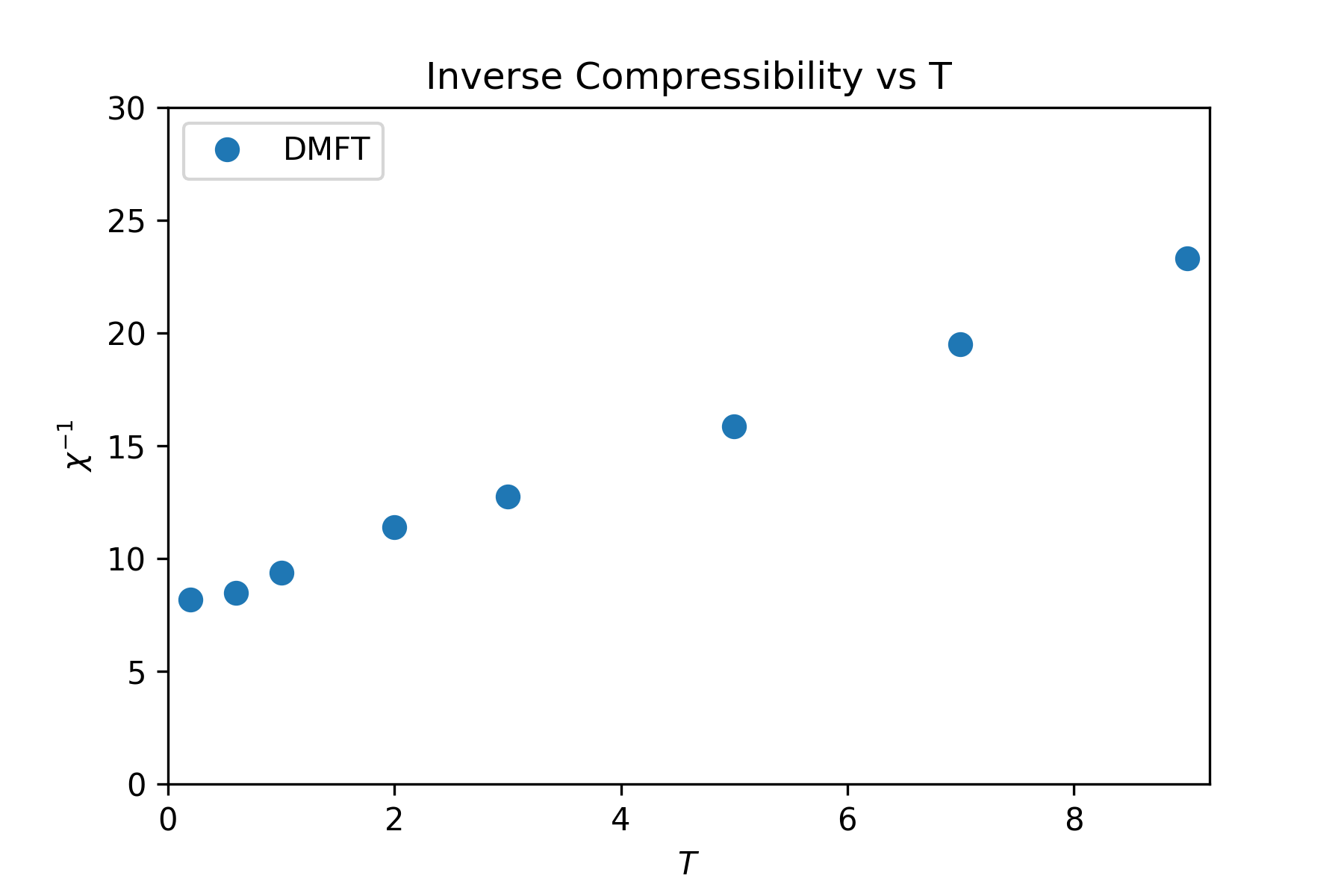}
\caption{Inverse compressibility $\chi^{-1}$ at $U=7.5$, $n=0.825$. \label{fig:compressibility}}
\end{figure}

Recent works have proposed looking for insight into $T$-linear resistivity by studying compressibility $\chi$ and diffusivity $\mathcal{D}$, related to transport by the Nernst-Einstein relation $\sigma_{\text{DC}}=\chi\mathcal{D}$, at intermediate and high temperatures.~\cite{hartnoll, devereaux, calandra2003}.
We obtain compressibility in DMFT using continuous-time interaction-expansion (CTINT) QMC as the impurity solver. We computed the DMFT solution for multiple values of $\m$ near the value corresponding to $n=0.825$ and took the numerical derivative to find $\chi^{-1}=\frac{d\m}{dn}$. No analytical continuation is necessary as $n=-2G(\t=\b^-)$. We present the data from DMFT in Fig.~\ref{fig:compressibility}.

To gain further insight, we summarize analytical results in the perturbative-hopping regime $t\ll T, U$, discussed in Refs.~\cite{mukerjee2007, hightemp}. In this limit, the chemical potential $\m$ and DC conductivity $\sigma_{\rm DC}$ can be calculated from the atomic Hubbard model, i.e., Hubbard model with $t\ra 0$. These expressions are
\beq
e^{\b\m}=&\frac{\sqrt{\d^2+(1-\d^2)y}-\d}{(1+\d)y}\equiv x\\
\sigma_{\rm DC}=&\frac{t}{4T}\(\frac{(1+\d)^2}{x+1/x+2}+\frac{(1-\d)^2}{xy+1/xy+2}\),
\eeq
where $\d=1-n$, $y=e^{-U/T}$ and the second expression is valid up to dimensionless constants.

From these expressions, we can take further limits to study the compressibility $\chi$, DC conductivity $ \sigma_{\rm DC}$, and diffusivity $\mathcal{D}\equiv\sigma_{\rm DC}/\chi$ in the intermediate-temperature $t\ll T\ll U$ or high-temperature $t\ll U \ll T$ limits. In the intermediate-temperature regime, we find
\beq
\mu_{\rm int}=& T\log\frac{n}{2(1-n)}\\
\chi_{\rm int}^{-1}=&\frac{T}{n(1-n)}\\
\rho_{\rm DC, int}=&\frac{2T}{n(1-n)}\\
\mathcal{D}_{\rm int}^{-1}=&2.
\eeq
On the other hand, in the high-temperature regime, we find
\beq
\mu_{\rm high}=& T\log\frac{n}{2-n}\\
\chi_{\rm high}^{-1}=&\frac{2T}{n(2-n)}+\frac{U}{2}\\
\rho_{\rm DC, high}=&\frac{8T}{1-(1-n)^4}\\
\mathcal{D}_{\rm high}^{-1}=&\frac{4}{1+(1-n)^2}.
\eeq
We remark that $\mathcal{D}$ saturates to different expressions in both regimes. As noted in the main text, the crossover between the two expressions is set by the change in the temperature-dependence of the chemical potential, which follows from setting  $y\ll \frac{\d^2}{1-\d^2}$. At small doping, the crossover temperature is therefore suppressed from the na\"ive $\sim U$ to $\sim U/\log((1-\d^2)/\d^2)$.

\end{document}